\documentclass[prl,twocolumn,superscriptaddress,nofootinbib]{revtex4-2}

\usepackage{preamble}

\begin{document}

\title{Nonperturbative optomechanics with thin, weakly reflective intracavity dielectrics}

\author{Andrew Laeuger\,\orcidlink{0000-0002-8212-6496}}
\email{alaeuger@caltech.edu}
\affiliation{Department of Physics and Astronomy, Caltech, Pasadena, California 91125, USA}

\author{Nancy Aggarwal\,\orcidlink{0000-0002-0026-3877}}
\email{nqaggarwal@ucdavis.edu}
\affiliation{Department of Physics and Astronomy, University of California, Davis, California 95616, USA}

\date{\today}

\begin{abstract}
Levitated dielectric sensors are typically treated as perturbative probes of a pre-existing optical field. We show that a plate capturing the full optical mode can strongly restructure the cavity field, even at modest refractive indices and subwavelength thicknesses. By partitioning the cavity into coupled subcavities, the plate creates new optical resonances and trapping sites. Perturbative predictions for the cavity optomechanical coupling can fail at thicknesses as small as $\sim\lambda/(10n)$, with cavity locking suppressing the coupling by roughly three orders of magnitude. Meanwhile, the subcavity structure enables appreciable probe-beam cooling on resonance, which is absent for a purely dispersive single mode coupling. These results establish a regime in which the levitated sensor helps define the cavity field rather than merely probing it.
\end{abstract}

\maketitle

\ssec{Introduction}
Optical cavities coupled to mechanical degrees of freedom provide a powerful platform for precision measurements, relying on the sensitivity of resonant optical fields to small mechanical motions and on the ability of radiation pressure to manipulate those degrees of freedom. Cavity optomechanics has enabled studies of radiation-pressure backaction and cooling, quantum optomechanics, squeezed-light generation, and gravitational-wave detection \cite{CavityOptomechanics,Kippenburg2007, Li2021,cripe2019measurement, Metcalfe2014,Cripe:2017cvb, Marquardt2009,aggarwal2020room}. Optomechanical resonators have also emerged as exceptionally sensitive probes of weak forces, with applications in force, torque, and acceleration sensing \cite{Gavartin2012, Ranjit:2016dlq, Liang2023, Guha2020, Komori2020, Zhou2021, Ahn:2019cbu, Ju:2023egl}. In these experiments the mechanical element is often treated as a probe of an optical field that it only weakly modifies.

In levitated optomechanics, the sensor is positioned by optical forces, but is likewise often modeled as a perturbative polarizable particle. Its position-dependent interaction with an optical field determined independently of the sensor defines the optomechanical coupling \cite{ChangSpheres,ChangThinDiscs, Millen2020}. This picture is natural for nanoscale particles, which interact primarily with the local field intensity, but also extends to larger objects treated using their full optical scattering response \cite{Maurer:2021zgx, Maurer:2022nvj, Reigue:2022thv}. The key distinction is whether the object's presence and motion produce only a small modification of the optical field. In the single-mode version of this picture, the sign of the optical damping is tied to the sign of the detuning \cite{CavityOptomechanics,Marquardt2007}.

Other geometries, notably membrane-in-the-middle and membrane-at-the-edge, use dielectrics that substantially modify the cavity spectrum, producing position-dependent frequency shifts, mode splitting, and avoided crossings \cite{Thompson:2007hno,jayich2008dispersive,Dumont:2019ziy,stambaugh2015membrane,xuereb2010optomechanical,xuereb2012dynamical,DumontThesis}. Even with the cavity modes and frequencies heavily altered, the extensively studied regime of thin, high reflectivity dielectric membranes still admits a perturbative treatment of the motion:  the optomechanical coupling is obtained by linearizing around the modified cavity eigenmodes and eigenfrequencies \cite{Thompson:2007hno,jayich2008dispersive}, rather than by a self-consistent modification of the optical field by the displaced dielectric.

Here we present a qualitatively different regime: a plate-like dielectric sensor, levitated by the optical forces, whose transverse extent is larger than the optical beam waist such that it captures essentially the entire optical mode \cite{Muller:2015idf}. Such sensors are motivated in part by proposals for detecting high-frequency gravitational waves \cite{LSD2012, LSD2020, AndrewsGRNote}, and high-aspect-ratio dielectric structures have recently been levitated experimentally \cite{LSD2022}. We focus on the sensor's on-axis dynamics, leaving detailed analyses of transverse motion to future work.

\begin{figure}[t]
    \centering
    \hspace*{0.6cm}\includegraphics[width=0.8\linewidth]{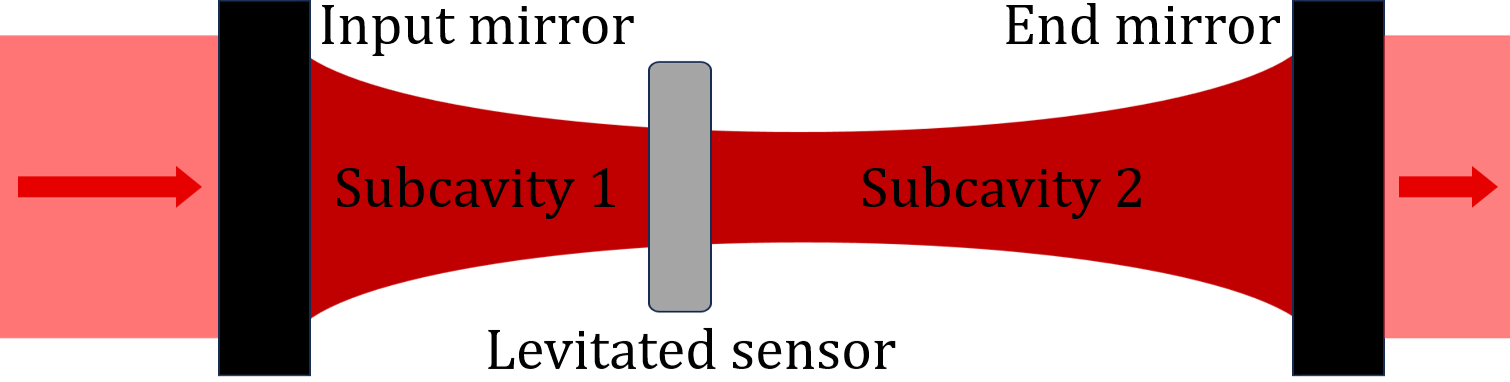}
    \vspace{-0.5cm}
    \includegraphics[width=0.9\linewidth]{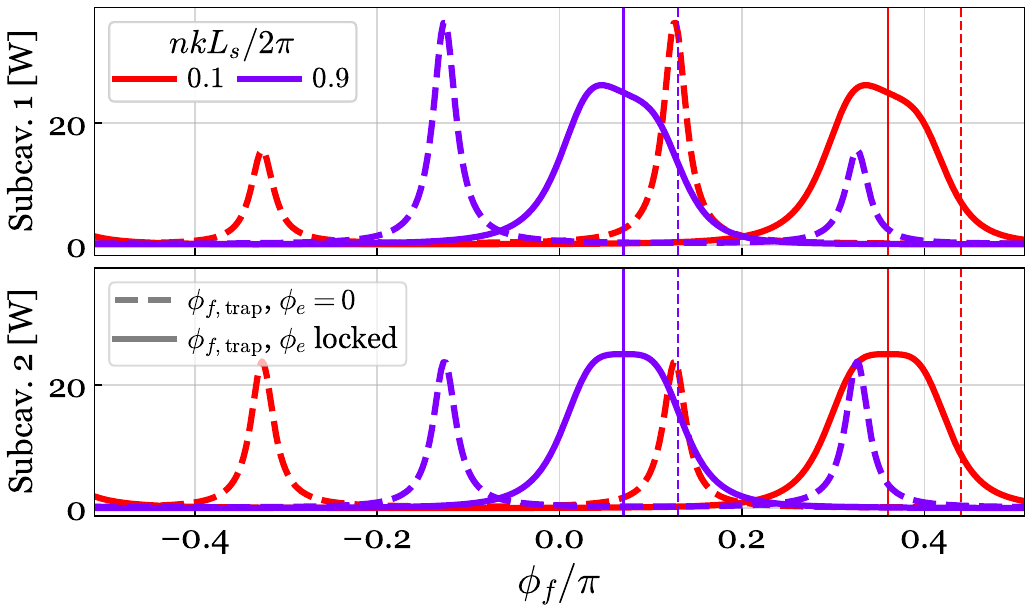}
    \caption{\textit{Top:} The experimental layout considered in this paper. \textit{Bottom:} The power in Subcavities 1 and 2 versus $\phi_f$ for two choices of $L_s$, where $\phi_e$ is either 0 or its locked value. When $\phi_e=0$, two separate resonances are distinguished by their Lorentzian structure. Locking the cavity makes the two separate Lorentzians converge into a single unified resonance. Vertical lines appear at the stable trap sites.}
    \label{fig:subcavitylocked}
\end{figure}

The central finding is that an extended dielectric sensor can require a fully non-perturbative treatment even when it is only moderately reflective. The intracavity field, optical force, and equilibrium position can be substantially reorganized well before the dielectric becomes optically thick. Consequently, the sensor position cannot be treated simply as a parameter in a pre-existing cavity mode; the optical and mechanical responses must be determined self-consistently. We find that the plate creates coupled-subcavity resonances and thickness-dependent trapping sites, while conventional perturbative expressions for the cavity frequency shift and optomechanical coupling can fail even for $L_s\ll\lambda$. Cavity locking can strongly suppress the coupling while increasing the trap stiffness. The modified cavity structure also permits cooling on resonance, which is not possible with purely dispersive single-mode coupling.

In this letter, we investigate these effects using an exact one-dimensional field solution and a linearized two-photon description of carrier and sideband propagation. Unlike modal treatments of membranes with varying reflectivities \cite{jayich2008dispersive,Muller:2015idf}, our approach requires no truncation of optical modes: the dielectric enters directly into the boundary conditions for both carrier and sidebands. The mathematical construction is presented in Ref. \cite{FormalismPaper}. These phenomena are not restricted to freely levitated sensors and can also arise for membranes of similar geometry. In a membrane-based setup, however, the equilibrium position depends nontrivially on the relative optical and mechanical spring constants (see the Supplemental Material \cite{Supplemental}), making the quantitative behavior more parameter-dependent.

\ssec{Model and exact field solution}
In Figure \ref{fig:subcavitylocked}, we consider a high intensity trap beam entering a Fabry-P\'erot cavity; this laser generates the optical force which levitates the planar dielectric sensor. Additional lower intensity coaxial beams can be employed for positional readout of the sensor, for example. Regardless of the exact spatial mode structure inside the cavity, we assume that the sensor cross section well exceeds the beam waist where it is located, so that scattering occurs primarily along the cavity axis -- an assumption of plane wave propagation thus sufficiently captures the relevant physics. Notably, sensors with thickness equal to integer multiples of $\lambda/2n$ are transparent in this scheme; such geometries must receive a full three-dimensional treatment and thus lie outside the scope of this work.

For our numerical results, we use a cavity of length $L_c=1$ m and finesse $\mathcal{F}\approx 77$, laser wavelength $\lambda=2\pi c/\omega_0=1550$ nm, input trap power $P=1$ W, beam waist $w_0=37.5~\mu$m, and a silica ($n=1.45$) sensor of radius $R=75~\mu$m and thickness $L_s$. The macroscopic distance from the input mirror to the sensor is $L_t$. All optical elements are taken to be lossless.

The essential modification to standard levitated-particle calculations is that the dielectric is included directly in the optical boundary conditions. We represent the carrier and sideband electric fields by two-photon quadrature vectors. Free-space propagation rotates the quadrature vectors by the accumulated phase of the carrier over that length, with the sidebands accruing an additional frequency-dependent complex rotation \cite{CavesAndSchumaker1, CavesAndSchumaker2, CorbittFramework}. At the dielectric interfaces, the fields are related by the usual Fresnel coefficients, while propagation through the finite thickness includes the phase accumulated inside the dielectric. Solving the resulting linear field equations gives the carrier fields at every location in the cavity for arbitrary sensor position and thickness.
The DC and AC forces on the sensor are obtained by integrating the Lorentz force density over the polarized dielectric \cite{ForceLetter}, using the same carrier and sideband fields self-consistently determined from the optical propagation equations.

\ssec{Creation of coupled subcavities}
We first examine the steady-state trap field as a function of sensor thickness and position. Figure~\ref{fig:subcavitylocked} shows the circulating powers on the two faces, $\phi_f$ denotes the sensor position -- specifically, the phase of the front face of the sensor relative to the trap beam carrier field -- while $\phi_e$ is half the phase accumulated by the carrier in a round trip of the cavity, modulo $2\pi$. We first set $\phi_e=0$, as in the limit of a vanishingly reflective sensor, this value would keep the full cavity on resonance. The circulating powers take the form of pair of Lorentzians whose widths are set primarily by the cavity finesse, indicating that the insertion of the sensor generates new optical resonances.

These resonances emerge because reflection from the planar sensor breaks the full cavity into a sequence of coupled subcavities: Subcavity 1, formed by the input mirror and the left face of the sensor, and Subcavity 2, formed by the right face and the end mirror. At each optical resonance, one subcavity achieves a large internal power which partially leaks over into the other cavity, creating a local maximum in that power as well.

The resonance condition is not equivalent to simply maintaining an integer number of wavelengths in either subcavity. Because the sensor has only a mild reflectivity coefficient $\rho_s=(n-1)/(n+1)\simeq0.18$, the fields in the subcavities on either side of the dielectric remain substantially coupled to one another. Each resonance is ``led'' by the subcavity which achieves the larger steady-state power, and the coupling between them produces the thickness-dependent locations of the resonances and trap sites. The resonance led by Subcavity 1 achieves much greater power in that subcavity due to constructive interference between the fields entering from the input mirror and from Subcavity 2.
This cavity-splitting phenomenon is familiar from intracavity membrane systems \cite{Thompson:2007hno,jayich2008dispersive,Xu:2013hgm,xu2015quantum,stambaugh2015membrane,Dumont:2019ziy,DumontThesis,Yao:2024tpc,deJong2020}; we explore this connection further in the Supplemental Material.

In Figure \ref{fig:subcavitylocked}, we also note the stable sensor trap sites, which are determined by the simultaneous conditions
\begin{equation}
F_{\rm DC}(\phi_f)=0,
\qquad
\partial_{\phi_f}F_{\rm DC}<0.
\label{eq:trap}
\end{equation}
When $\phi_e=0$, the radiation pressure incident from either side disagrees at the resonant $\phi_f$, so the equilibria lie in between the peaks, where the circulating power is small. 

\ssec{Breakdown of perturbative optomechanics}
We directly test the departure from the perturbative regime via the cavity resonance frequency and optomechanical coupling. In conventional levitated optomechanics, the lowest-order perturbative shift to the cavity resonant frequency $\omega_c$ from inserting a dielectric is \cite{ChangSpheres, Joannopoulos_Johnson_Winn_Meade_2008}
\begin{equation}
\frac{\delta\omega_c}{\omega_c}
\approx
-\frac12
\frac{\int d^3r\,\delta\epsilon(\vec r)\,\vec{\mathbb E}_0^2(\vec r)}
{\int d^3r\,\vec{\mathbb E}_0^2(\vec r)},
\label{eq:pertshift}
\end{equation}
where $\vec{\mathbb E}_0$ is the unperturbed cavity field. For a cavity with nodes at its boundaries, the corresponding lowest-order optomechanical coupling $g_0\equiv k^{-1}\partial\omega_c/\partial x$ is
\begin{equation}
 g_0\approx-
 \frac{n^2-1}{kL_c}
 \sin(kL_s)\sin(2\phi_f+kL_s)\omega_c.
\label{eq:g0pert}
\end{equation}

\begin{figure}[t]
    \centering
    \includegraphics[width=\linewidth]{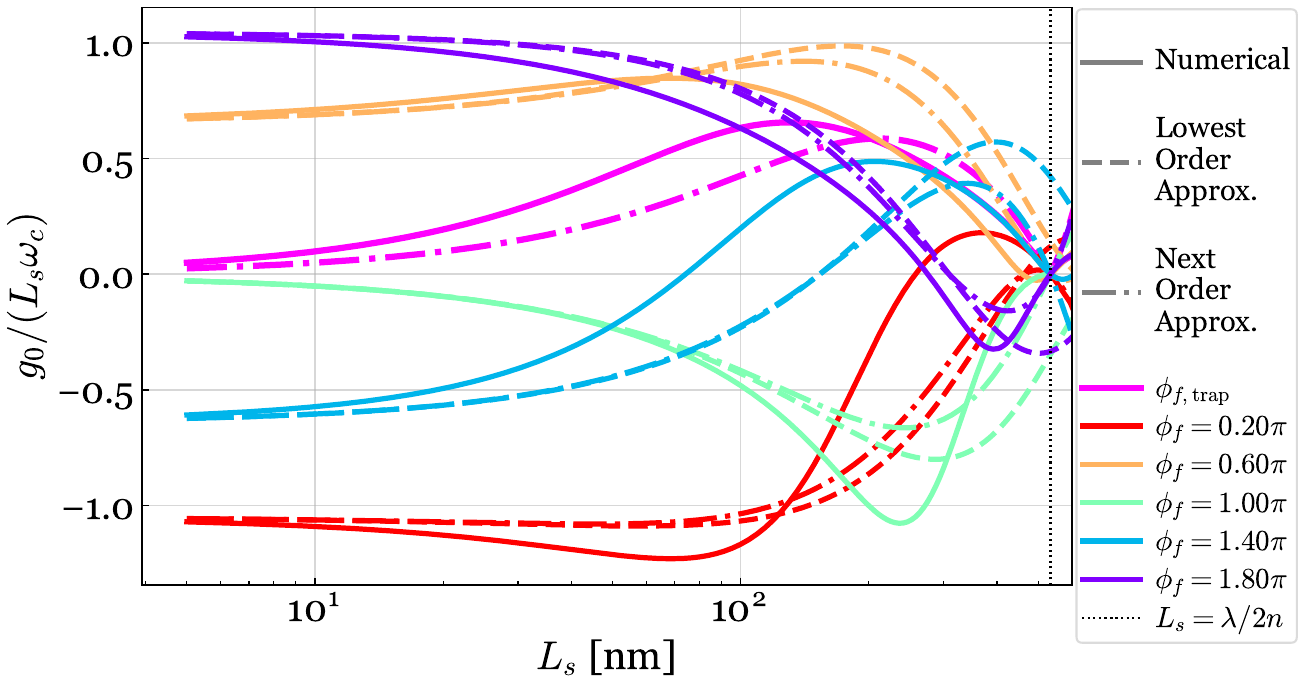}
    \caption{The optomechanical coupling $g_0$ versus $L_s$ for several sensor positions. Solid curves are obtained by maximizing the transmitted power in the exact cavity solution; dashed and dot-dashed curves use the lowest- (Eq. \eqref{eq:g0pert}) and next-to-lowest-order perturbative expressions. The perturbative results depart from the exact solution by $L_s\sim\lambda/(10n)$.}
    \label{fig:g0}
\end{figure}

Figure~\ref{fig:g0} compares the lowest-order and next-to-lowest-order \cite{Joannopoulos_Johnson_Winn_Meade_2008} perturbative expressions for $g_0$ with the exact cavity solution, where $\omega_c$ is defined by maximizing the transmitted power. Proper comparison to these perturbative expressions requires keeping the cavity length fixed when inserting the dielectric, which is done by choosing $\phi_e=(n-1)kL_s$. For very small $L_s$, the perturbative integrals agree well with the exact result. By $L_s\sim\lambda/(10n)$, however, both perturbative approximations have departed visibly; including the first field correction provides little improvement. In the Supplemental Material, we observe a similar departure from perturbative results for the resonant frequency shift $\delta\omega_c$.

Thus, the breakdown occurs while the geometric thickness is still much smaller than the optical wavelength. The relevant small parameter is not simply $L_s/\lambda$, but whether the dielectric-induced reflection and the resulting multiple reflections are negligible compared with the cavity response. A sensor with a modest index of refraction can therefore produce strong changes to the electric field throughout the entire cavity even when $L_s\ll\lambda$.
Related work has shown that systems with highly reflective membranes can be modeled using a few optical modes whereas weakly reflective membranes are strong perturbers: perturbative expansions in empty-cavity modes can require hundreds of longitudinal modes without fully reproducing the exact one-dimensional result \cite{jayich2008dispersive,Muller:2015idf}.
Here, we have directly quantified the breakdown of the standard perturbative expressions for the cavity-frequency shift and optomechanical coupling and shown that it occurs even when the dielectric thickness remains much smaller than the optical wavelength.

\ssec{Impact of cavity locking}
The cavity field restructuring has direct dynamical consequences, too: locking the cavity substantially modifies the mechanical and optical dynamics of the system. Here, the ``locked" cavity configuration is the pair $(\phi_f,\phi_e)$ which solves Eq. \eqref{eq:trap} while maximizing the power transmitted through the end mirror. That is, we adjust the cavity length until $dP_\text{trans}/d\phi_e=0$ at the stable trap site. More details appear in the Supplemental Material.

\begin{figure}[t]
    \centering
    \includegraphics[width=\linewidth]{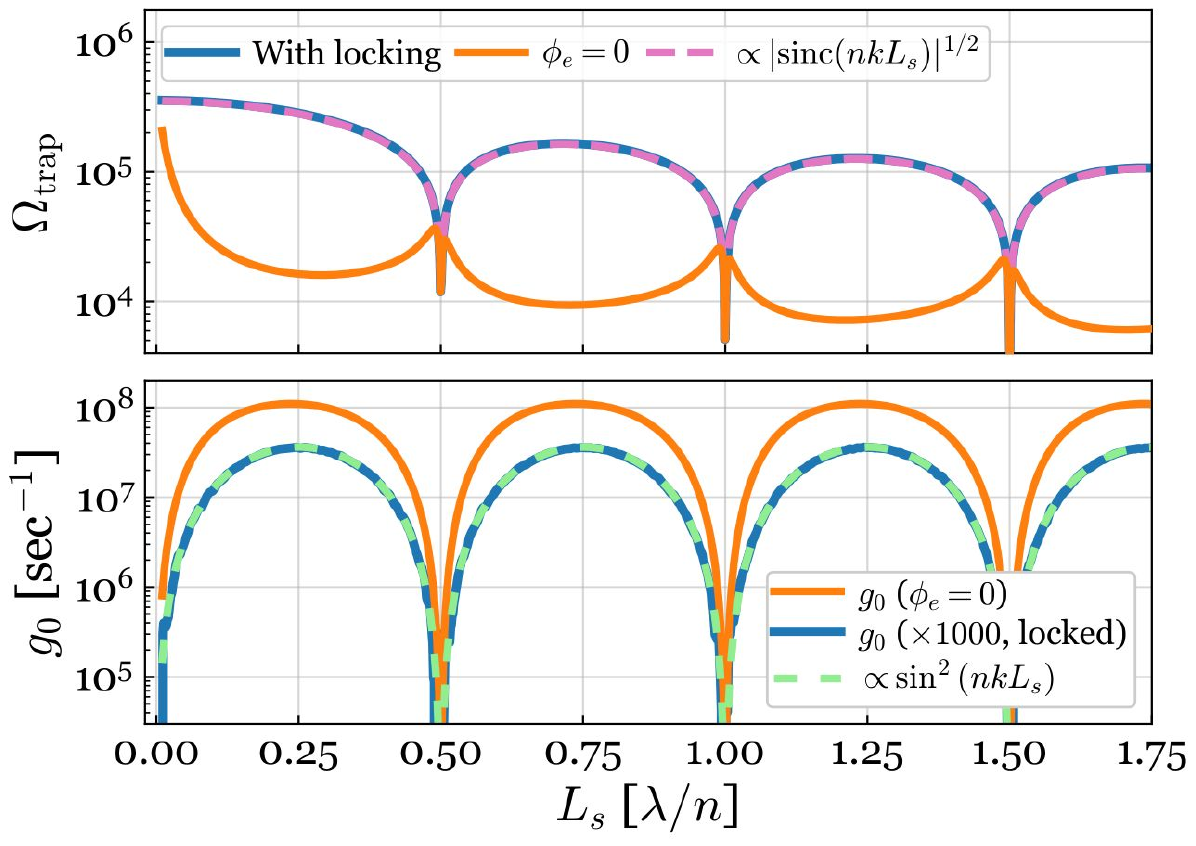}
    \caption{The optical trap frequency $\Omega_\text{trap}$ and optomechanical coupling $g_0$ versus $L_s$ for an unlocked cavity ($\phi_e=0$) and a cavity locked to maximal transmitted power. For the parameters used here, locking reduces $g_0$ by a factor of $\sim3000$ and increases $\Omega_\text{trap}$ by up to a factor of 20. Phenomenological fits for the locked quantities appear as dashed lines.}
    \label{fig:lockedg0}
\end{figure}

Figure~\ref{fig:lockedg0} compares the trap resonant frequency $\Omega_\text{trap}\equiv\sqrt{-k(\partial_{\phi_f}F_\text{DC})/m}$ and optomechanical coupling with and without this locking condition. Both quantities demonstrate periodicity in the dielectric half-wavelength $\lambda/(2n)$. This periodicity follows directly from the fact that the fields, and hence the forces, are periodic when the dielectric accumulates an additional $2\pi$ round-trip phase. The trap frequency also exhibits a secular $L_s^{-1/2}$ decrease through the $\propto L_s$ scaling of the sensor mass.

Notably, for the representative parameters used here, locking reduces $g_0$ by approximately a factor of $3000$ while increasing $\Omega_\text{trap}$ by a factor of up to 20. The origin of the suppression and amplification, respectively, can be understood from the subcavity structure. Without locking, the sensor sits between the two resonant features in the transmitted power, where the transmitted power has a small but nonzero derivative with respect to sensor position. Under the locking condition, however, the two relevant resonances in each subcavity nearly converge at the equilibrium configuration, leaving a wider, non-Lorentzian merged resonance structure -- see Figure \ref{fig:subcavitylocked}.

The radiation pressure on the sensor from either side now balances within the merged resonance, rather than in the low power region between the two distinct positional Lorentzians. The differential incident power therefore changes much more rapidly with sensor position near the equilibrium, generating a stiffer trap. Furthermore, the transmitted power becomes insensitive to sensor position at first order -- a result of impedance matching in the cavity, which makes the Lorentzians in Subcavity 2 equal in magnitude and the trap site fall at the middle of their merged structure. Although a displacement of the sensor and change in laser frequency are not completely interchangeable, both perturb the optical phase accumulated in the subcavities. The hierarchy of transmitted-power derivatives therefore provides useful intuition for the corresponding hierarchy in $g_0$.

Interestingly, the locked coupling is fit very well by a function proportional to $\sin^2(nkL_s)$. The same scaling appears in the power reflectivity of the complete dielectric, including internal etalon effects \cite{jayich2008dispersive, FormalismPaper}, suggesting that the residual coupling may be controlled by the weak net reflectivity of the finite slab. Similarly, the locked trap frequency scales like $|\text{sinc}(nkL_s)|^{-1/2}$, a form which appears related to the same physics driving the residual coupling magnitude. A more complete analytic treatment of these scaling relations is left for future work.

Although the lineshape of power versus sensor position is distinctly non-Lorentzian in the locked configuration, the lineshape of power versus detuning frequency remains Lorentzian. Since our locking procedure maximizes the transmitted power, a detuning frequency of $\Delta=0$ denotes the secondary beam being tuned to maximal transmitted power. Interestingly, the system linewidth $\kappa$, given by the detuning angular frequency which halves the transmitted power of the locked cavity, is not only $L_s$-independent with the cavity locked, but it is also equal to the empty-cavity value, $\pi c/2L\mathcal{F}$.

\ssec{On-resonance cooling}
\begin{figure}[t]
    \centering
    \includegraphics[width=\linewidth]{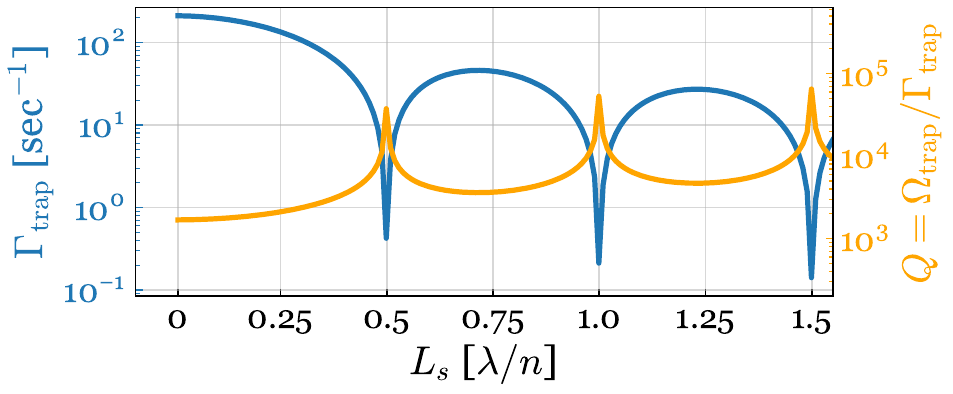}
    \caption{The trap beam induced cooling linewidth $\Gamma_\mathrm{trap}$ and resulting trap $Q$-factor, with the cavity in its locked configuration. Significant cooling appears on resonance because the locking merges the distinct single mode resonances into a joint multimode feature in the vicinity of the trap.}
    \label{fig:cooling}
\end{figure}
This subcavity structure also changes dynamical backaction. In typical optomechanical setups, a weaker, detuned secondary beam cools the sensor's thermal motion. For conventional single-mode coupling, the sign of optical damping is tied to detuning \cite{CavityOptomechanics,Marquardt2007,Schliesser2006,Dantan:2007bqd,Arcizet:2006wlz,Cohadon:1999xf,Gigan:2006xjb}, so a zero-detuned beam produces neither cooling nor heating. On-resonance cooling has nevertheless been predicted for highly reflective membranes in the two-mode regime, where sideband scattering interferes between optical eigenmodes at the bare degeneracy point between the split resonances \cite{jayich2008dispersive}. On-resonance cooling has also been observed for generalized coupling, where a position-dependent cavity linewidth (dissipative coupling) interferes with the dispersive coupling of a single effective optical mode \cite{Sawadsky:2014bra}.

We test whether this effect occurs for moderately reflective dielectrics by calculating the trap-beam-induced damping with the cavity locked. Figure \ref{fig:cooling} shows the resulting linewidth and trap quality factor, $Q\equiv\Omega_\text{trap}/\Gamma_\text{trap}$. We find substantial cooling from the on-resonance trap beam; gas damping is five to six orders of magnitude weaker \cite{LSD2012, LSD2020}. Substantial cooling and a nonzero optical trap frequency (Figure \ref{fig:lockedg0}) both persist in the $L_s\rightarrow0$ limit because both the optical spring constant and mass scale with $L_s$ in this regime.

This unusual cooling on resonance follows from the locked subcavity powers in Figure \ref{fig:subcavitylocked}. Locking merges the two resonances initially present at $\phi_e=0$ into a single broad feature. As we have shown in relation to $g_0$, our locking condition strongly suppresses the resonant frequency shift, leaving transmitted power essentially independent of $\phi_f$ at equilibrium. Nevertheless, Subcavity 1 retains a finite negative power slope at $\phi_{f,\mathrm{trap}}$. Since the resonance does not move with $\phi_f$ at this point, a sensor displacement changes the power stored on the input side of the plate without shifting the resonance. Because this stored power cannot adjust instantaneously, the resulting optical force from Subcavity 1 lags the sensor motion, producing damping.

Unlike previously studied on-resonance cooling mechanisms, here $\Delta=0$ places the laser at the maximum transmitted power of the occupied, locked cavity, and the single-mode sign rule is still violated. Moreover, the coupled-subcavity resonances converge at the mechanical equilibrium of the freely levitated sensor, rather than at a position chosen for a mounted membrane. The resulting power response is a merged, non-Lorentzian multimode feature whose asymmetry produces the corresponding sideband-force response. Notably, this implies nonzero cooling on a trapped sensor at $\Delta=0$ when the cavity is unlocked, too -- its magnitude is just weaker due to the smaller derivative of the optical force with respect to $\phi_f$.

The standard heating/cooling split can be seen by setting $\phi_e=0$ and holding the sensor at either distinct resonance in $\phi_f$. There, the sensor sits at the center of an effectively single-mode response, and the stored power on either side is insensitive to position to first order. A displacement shifts the resonance without changing the stored power at first order. Although the cavity response still lags the displacement, no first-order force modulation is produced, so the net cooling at $\Delta=0$ vanishes. This behavior, together with cooling from a weak secondary beam at generic detuning, is demonstrated in the Supplemental Material.

Figure \ref{fig:cooling} indicates that the trap beam alone can source substantial thermal cooling, with $Q$ in the range $10^3-10^4$ for 1 W input power and scaling as $\sim1/\sqrt{P}$. The zero-detuning effect is significant not only in its magnitude for this parameter choice, but also as a dynamical signature of sensor-induced cavity restructuring.

\ssec{Conclusion}
Our results demonstrate a regime of cavity optomechanics in which a laterally-extended dielectric cannot be treated as a perturbative probe of a pre-existing optical field. A plate-like sensor with modest reflectivity creates coupled subcavities and new optical resonances, while its own equilibrium position is determined by the radiation pressure generated by those reorganized fields. The resulting nonperturbative behavior can be observed even for thicknesses  $L_s\ll\lambda$, though the precise $L_s$ at which the physics departs from the perturbative regime depends nontrivially on the sensor reflectivity, cavity finesse, and detuning angles. Furthemore, the underlying optical restructuring does not require free levitation and can also arise for mechanically supported membranes; levitation can be thought of here as the compliant-membrane limit of the self-consistent optical problem.

Upon entering this nonperturbative sector, the trap position and frequency acquire a nontrivial thickness dependence, and conventional perturbative expressions for the cavity-frequency and optomechanical couplings deviate from exact values. These effects arise from a common physical origin: the sensor changes the global optical boundary conditions rather than merely sampling a preconstructed internal field. The same restructured subcavity fields also result in suppression of coupling in presence of locking and allow cooling on resonance.

We have employed a simplified one-dimensional propagation model appropriate when the dielectric captures the full transverse optical mode. Because the sensor radius well exceeds the beam waist size, we expect the qualitative nonperturbative behavior reported here to persist under a fully three-dimensional treatment. The exact field and force formalism can be extended to accommodate such a treatment, along with more complicated optical layouts and multilayer sensors. The predicted departure from perturbative behavior around $\sim\lambda/(10n)$ is directly testable: measuring trap frequency and optomechanical coupling  via cavity transmission as a function of sensor thickness at fixed refractive index would provide a direct experimental probe of the nonperturbative predictions in Figure~\ref{fig:g0}. For proposed levitated-sensor based gravitational-wave detectors, our self-consistent treatment provides the starting point for calculating signal and noise transfer functions and for systematically exploring detector design space, topics which are examined thoroughly in our accompanying paper \cite{FormalismPaper}.

\ssec{Acknowledgments}
A.L. is grateful for support from the Fannie and John Hertz Foundation through a Hertz Fellowship. N.A. is supported by the UC Noyce Initiative and the Department of Energy award number DE-SC0009999. We also thank Rana Adhikari, Dan Brown, Yanbei Chen, Su Direkci, Vincent Dumont, Shafaq Elahi, James Gardner, Andrew Geraci, Kevin Kuns, Shane Larson, Aaron Markowitz, Jack Sankey, Vivishek Sudhir, Dalziel Wilson, and Christopher Wipf for fruitful discussions throughout this work.

\bibliography{main}

@article{Cripe:2017cvb,
    author = "Cripe, Jonathan and Aggarwal, Nancy and Singh, Robinjeet and Lanza, Robert and Libson, Adam and Yap, Min Jet and Cole, Garrett D. and McClelland, David E. and Mavalvala, Nergis and Corbitt, Thomas",
    title = "{Radiation-Pressure-Mediated Control of an Optomechanical Cavity}",
    eprint = "1710.04700",
    archivePrefix = "arXiv",
    primaryClass = "quant-ph",
    doi = "10.1103/PhysRevA.97.013827",
    journal = "Phys. Rev. A",
    volume = "97",
    number = "1",
    pages = "013827",
    year = "2018"
}

@article{cripe2019measurement,
  title={Measurement of quantum back action in the audio band at room temperature},
  author={Cripe, Jonathan and Aggarwal, Nancy and Lanza, Robert and Libson, Adam and Singh, Robinjeet and Heu, Paula and Follman, David and Cole, Garrett D and Mavalvala, Nergis and Corbitt, Thomas},
  journal={Nature},
  volume={568},
  number={7752},
  pages={364--367},
  year={2019},
  publisher={Nature Publishing Group UK London}
}

@article{aggarwal2020room,
  title={Room-temperature optomechanical squeezing},
  author={Aggarwal, Nancy and Cullen, Torrey J and Cripe, Jonathan and Cole, Garrett D and Lanza, Robert and Libson, Adam and Follman, David and Heu, Paula and Corbitt, Thomas and Mavalvala, Nergis},
  journal={Nature Physics},
  volume={16},
  number={7},
  pages={784--788},
  year={2020},
  publisher={Nature Publishing Group UK London}
}

@article{LSD2020,
    author = "Aggarwal, Nancy and Winstone, George P. and Teo, Mae and Baryakhtar, Masha and Larson, Shane L. and Kalogera, Vicky and Geraci, Andrew A.",
    title = "{Searching for New Physics with a Levitated-Sensor-Based Gravitational-Wave Detector}",
    eprint = "2010.13157",
    archivePrefix = "arXiv",
    primaryClass = "gr-qc",
    doi = "10.1103/PhysRevLett.128.111101",
    journal = "Phys. Rev. Lett.",
    volume = "128",
    number = "11",
    pages = "111101",
    year = "2022"
}

@article{LSD2022,
    author = "Winstone, George and others",
    collaboration = "LSD",
    title = "{Optical Trapping of High-Aspect-Ratio NaYF Hexagonal Prisms for kHz-MHz Gravitational Wave Detectors}",
    eprint = "2204.10843",
    archivePrefix = "arXiv",
    primaryClass = "physics.optics",
    doi = "10.1103/PhysRevLett.129.053604",
    journal = "Phys. Rev. Lett.",
    volume = "129",
    number = "5",
    pages = "053604",
    year = "2022"
}

@article{CorbittFramework,
    author = "Corbitt, Thomas and Chen, Yanbei and Mavalvala, Nergis",
    title = "{Mathematical framework for simulation of quantum fields in complex interferometers using the two-photon formalism}",
    eprint = "quant-ph/0502088",
    archivePrefix = "arXiv",
    reportNumber = "LIGO-P030070-00-R, AEI-2005-006",
    doi = "10.1103/PhysRevA.72.013818",
    journal = "Phys. Rev. A",
    volume = "72",
    pages = "013818",
    year = "2005"
}

@article{LSD2012,
    author = "Arvanitaki, Asimina and Geraci, Andrew A.",
    title = "{Detecting high-frequency gravitational waves with optically-levitated sensors}",
    eprint = "1207.5320",
    archivePrefix = "arXiv",
    primaryClass = "gr-qc",
    doi = "10.1103/PhysRevLett.110.071105",
    journal = "Phys. Rev. Lett.",
    volume = "110",
    number = "7",
    pages = "071105",
    year = "2013"
}

@article{ChangThinDiscs,
       author = {{Chang}, D.~E. and {Ni}, K. -K. and {Painter}, O. and {Kimble}, H.~J.},
        title = "{Ultrahigh-Q mechanical oscillators through optical trapping}",
      journal = {arXiv e-prints},
         year = 2010,
        month = dec,
          eid = {arXiv:1101.0146},
        pages = {arXiv:1101.0146},
          doi = {10.48550/arXiv.1101.0146},
archivePrefix = {arXiv},
       eprint = {1101.0146},
 primaryClass = {quant-ph},
       adsurl = {https://ui.adsabs.harvard.edu/abs/2011arXiv1101.0146C}
}

@ARTICLE{ChangSpheres,
       author = {{Chang}, D.~E. and {Regal}, C.~A. and {Papp}, S.~B. and {Wilson}, D.~J. and {Ye}, J. and {Painter}, O. and {Kimble}, H.~J. and {Zoller}, P.},
        title = "{Cavity opto-mechanics using an optically levitated nanosphere}",
      journal = {Proceedings of the National Academy of Science},
         year = 2010,
        month = jan,
       volume = {107},
       number = {3},
        pages = {1005-1010},
          doi = {10.1073/pnas.0912969107},
archivePrefix = {arXiv},
       eprint = {0909.1548},
 primaryClass = {quant-ph},
       adsurl = {https://ui.adsabs.harvard.edu/abs/2010PNAS..107.1005C}
}

@phdthesis{DumontThesis,
    author = "Dumont, Vincent",
    title = "{Quantum noise and mechanical membranes in microcavities}",
    year = "2022"
}

@article{CavesAndSchumaker1,
    author = "Caves, Carlton M. and Schumaker, Bonny L.",
    title = "{New formalism for two-photon quantum optics. 1. Quadrature phases and squeezed states}",
    doi = "10.1103/PhysRevA.31.3068",
    journal = "Phys. Rev. A",
    volume = "31",
    pages = "3068--3092",
    year = "1985"
}

@article{CavesAndSchumaker2,
    author = "Schumaker, Bonny L. and Caves, Carlton M.",
    title = "{New formalism for two-photon quantum optics. 2. Mathematical foundation and compact notation}",
    doi = "10.1103/PhysRevA.31.3093",
    journal = "Phys. Rev. A",
    volume = "31",
    pages = "3093--3111",
    year = "1985"
}

@article{CavityOptomechanics,
    author = "Aspelmeyer, Markus and Kippenberg, Tobias J. and Marquardt, Florian",
    title = "{Cavity Optomechanics}",
    eprint = "1303.0733",
    archivePrefix = "arXiv",
    primaryClass = "cond-mat.mes-hall",
    doi = "10.1103/RevModPhys.86.1391",
    journal = "Rev. Mod. Phys.",
    volume = "86",
    pages = "1391",
    year = "2014"
}

@book{Joannopoulos_Johnson_Winn_Meade_2008, 
    place={Princeton, NJ}, edition={2}, 
    title={Photonic crystals: Molding the flow of light}, 
    publisher={Princeton University Press}, 
    author={Joannopoulos, John D. and Johnson, Steven G. and Winn, Joshua N. and Meade, Robert D.}, 
    year={2008}}

@article{Schliesser2006,
       author = {{Schliesser}, A. and {Del'Haye}, P. and {Nooshi}, N. and {Vahala}, K.~J. and {Kippenberg}, T.~J.},
        title = "{Radiation Pressure Cooling of a Micromechanical Oscillator Using Dynamical Backaction}",
      journal = {\prl},
         year = 2006,
        month = dec,
       volume = {97},
       number = {24},
          eid = {243905},
        pages = {243905},
          doi = {10.1103/PhysRevLett.97.243905},
archivePrefix = {arXiv},
       eprint = {physics/0611235},
 primaryClass = {physics.optics}
}

@ARTICLE{Marquardt2007,
       author = {{Marquardt}, Florian and {Chen}, Joe P. and {Clerk}, A.~A. and {Girvin}, S.~M.},
        title = "{Quantum Theory of Cavity-Assisted Sideband Cooling of Mechanical Motion}",
      journal = {\prl},
         year = 2007,
        month = aug,
       volume = {99},
       number = {9},
          eid = {093902},
        pages = {093902},
          doi = {10.1103/PhysRevLett.99.093902},
archivePrefix = {arXiv},
       eprint = {cond-mat/0701416},
 primaryClass = {cond-mat.mes-hall}
}

@article{Dantan:2007bqd,
    author = "Dantan, Aurelien and Genes, Claudiu and Vitali, David and Pinard, Michel",
    title = "{Self-cooling of a movable mirror to the ground state using radiation pressure}",
    eprint = "0707.2038",
    archivePrefix = "arXiv",
    primaryClass = "quant-ph",
    doi = "10.1103/PhysRevA.77.011804",
    journal = "Phys. Rev. A",
    volume = "77",
    pages = "011804",
    year = "2008"
}

@article{Arcizet:2006wlz,
    author = "Arcizet, O. and Cohadon, P. -F. and Briant, T. and Pinard, M. and Heidmann, A.",
    title = "{Radiation-pressure cooling and optomechanical instability of a micro-mirror}",
    eprint = "quant-ph/0607205",
    archivePrefix = "arXiv",
    doi = "10.1038/nature05244",
    journal = "Nature",
    volume = "444",
    pages = "71",
    year = "2006"
}

@article{Cohadon:1999xf,
    author = "Cohadon, P. F. and Heidmann, A. and Pinard, M.",
    title = "{Cooling of a mirror by radiation pressure}",
    eprint = "quant-ph/9903094",
    archivePrefix = "arXiv",
    doi = "10.1103/PhysRevLett.83.3174",
    journal = "Phys. Rev. Lett.",
    volume = "83",
    pages = "3174--3177",
    year = "1999"
}

@article{Gigan:2006xjb,
    author = "Gigan, S. and Boehm, H. R. and Paternostro, M. and Blaser, F. and Langer, G. and Hertzberg, J. B. and Schwab, K. and Baeuerle, D. and Aspelmeyer, M. and Zeilinger, A.",
    title = "{Self-cooling of a micro-mirror by radiation pressure}",
    eprint = "quant-ph/0607068",
    archivePrefix = "arXiv",
    doi = "10.1038/nature05273",
    journal = "Nature",
    volume = "444",
    pages = "67--70",
    month = "11",
    year = "2006"
}

@article{Sawadsky:2014bra,
    author = "Sawadsky, Andreas and Kaufer, Henning and Nia, Ramon Moghadas and Tarabrin, Sergey P. and Khalili, Farid Ya. and Hammerer, Klemens and Schnabel, Roman",
    title = "{Observation of generalized optomechanical coupling and cooling on cavity resonance}",
    eprint = "1409.3398",
    archivePrefix = "arXiv",
    primaryClass = "quant-ph",
    doi = "10.1103/PhysRevLett.114.043601",
    journal = "Phys. Rev. Lett.",
    volume = "114",
    number = "4",
    pages = "043601",
    year = "2015"
}

@ARTICLE{AndrewsGRNote,
       author = {{Laeuger}, Andrew and {Elahi}, Shafaq Gulzar and {Klomp}, Shelby and {Larsen}, Jackson and {Sprague}, Jacob and {Wang}, Zhiyuan and {Winstone}, George and {Wroblewski}, Maddox and {Larson}, Shane L. and {Geraci}, Andrew A. and {Aggarwal}, Nancy},
        title = "{Gravitational wave signal and noise response of an optically levitated sensor in a Fabry-P{\'e}rot cavity}",
      journal = {\prd},
         year = 2026,
        month = sep,
       volume = {114},
       number = {6},
          eid = {064071},
        pages = {064071},
          doi = {10.1103/mhz8-h774},
archivePrefix = {arXiv},
       eprint = {2604.07302},
 primaryClass = {gr-qc},
}

@article{Dumont:2019ziy,
    author = "Dumont, Vincent and Bernard, Simon and Reinhardt, Christoph and Kato, Alex and Ruf, Maximilian and Sankey, Jack C.",
    journal = {Opt. Express},
    number = {18},
    pages = {25731--25748},
    publisher = {Optica Publishing Group},
    volume = {27},
    month = {Sep},
    year = {2019},
    title = "{Flexure-Tuned Membrane-at-the-Edge Optomechanical System}",
    url = {https://opg.optica.org/oe/abstract.cfm?URI=oe-27-18-25731},
    eprint = "1905.04594",
    archivePrefix = "arXiv",
    primaryClass = "quant-ph",
    doi = "10.1364/OE.27.025731",
}

@article{Thompson:2007hno,
    author = "Thompson, J. D. and Zwickl, B. M. and Jayich, A. M. and Marquardt, Florian and Girvin, S. M. and Harris, J. G. E.",
    title = "{Strong dispersive coupling of a high-finesse cavity to a micromechanical membrane}",
    eprint = "0707.1724",
    archivePrefix = "arXiv",
    primaryClass = "quant-ph",
    doi = "10.1038/nature06715",
    journal = "Nature",
    volume = "452",
    number = "7183",
    pages = "72--75",
    year = "2008"
}

@article{xuereb2012dynamical,
  title={Dynamical scattering models in optomechanics: going beyond the ‘coupled cavities’ model},
  author={Xuereb, Andr{\'e} and Domokos, Peter},
  journal={New Journal of Physics},
  volume={14},
  number={9},
  pages={095027},
  year={2012},
  publisher={IOP Publishing},
  doi = "10.1088/1367-2630/14/9/095027",
  
}

@article{xuereb2010optomechanical,
  title={Optomechanical cooling with generalized interferometers},
  author={Xuereb, Andr{\'e} and Freegarde, Tim and Horak, Peter and Domokos, Peter},
  journal={Physical review letters},
  volume={105},
  number={1},
  pages={013602},
  year={2010},
  publisher={APS},
  doi = "10.1103/PhysRevLett.105.013602",
  arxivPrefix = "arXiv",
  eprint = "1002.0463"

}

@article{stambaugh2015membrane,
  title={From membrane-in-the-middle to mirror-in-the-middle with a high-reflectivity sub-wavelength grating},
  author={Stambaugh, Corey and Xu, Haitan and Kemiktarak, Utku and Taylor, Jacob and Lawall, John},
  journal={Annalen der Physik},
  volume={527},
  number={1-2},
  pages={81--88},
  year={2015},
  publisher={Wiley Online Library},
  eprint = "1407.1709",
  archivePrefix = "arXiv",
  primaryClass = "physics.optics",
}

@article{xu2015quantum,
    author = "Xu, Xunnong and Gullans, Michael and Taylor, Jacob M.",
    title = "{Quantum Nonlinear Optics Near Optomechanical Instabilities}",
    eprint = "1404.3726",
    archivePrefix = "arXiv",
    primaryClass = "quant-ph",
    doi = "10.1103/PhysRevA.91.013818",
    journal = "Phys. Rev. A",
    volume = "91",
    pages = "013818",
    year = "2015"
}

@article{jayich2008dispersive,
    author = "Harris, J. G E. and Marquardt, F. and Clerk, A. A. and Girvin, S. M. and Thompson, J. D. and Yang, C. and Zwickl, B. M. and Sankey, J. C. and Jayich, A. M.",
    title = "{Dispersive optomechanics: a membrane inside a cavity}",
    eprint = "0805.3723",
    archivePrefix = "arXiv",
    primaryClass = "quant-ph",
    doi = "10.1088/1367-2630/10/9/095008",
    journal = "New J. Phys.",
    volume = "10",
    number = "9",
    pages = "095008",
    year = "2008"
}

@article{Xu:2013hgm,
    author = "Xu, Xunnong and Taylor, Jacob M.",
    title = "{Beating the standard quantum limit for force sensing with a coupled two-mode optomechanical system}",
    eprint = "1303.7469",
    archivePrefix = "arXiv",
    primaryClass = "quant-ph",
    doi = "10.1103/PhysRevA.90.043848",
    journal = "Phys. Rev. A",
    volume = "90",
    pages = "043848",
    year = "2014"
}

@ARTICLE{PDHlocking,
       author = {{Drever}, R.~W.~P. and {Hall}, J.~L. and {Kowalski}, F.~V. and {Hough}, J. and {Ford}, G.~M. and {Munley}, A.~J. and {Ward}, H.},
        title = "{Laser phase and frequency stabilization using an optical resonator}",
      journal = {Applied Physics B: Lasers and Optics},
         year = 1983,
        month = jun,
       volume = {31},
       number = {2},
        pages = {97-105},
          doi = {10.1007/BF00702605},
}

@article{Yao:2024tpc,
    author = {Yao, Xiong and de Jong, Matthijs H. J. and Li, Jie and Gr{\"o}blacher, Simon},
    title = "{Long-Range Optomechanical Interactions in SiN Membrane Arrays}",
    eprint = "2408.03066",
    archivePrefix = "arXiv",
    primaryClass = "physics.optics",
    doi = "10.1103/PhysRevX.15.011014",
    journal = "Phys. Rev. X",
    volume = "15",
    number = "1",
    pages = "011014",
    year = "2025"
}

@article{deJong2020,
author = {Matthijs H. J. de Jong and Jie Li and Claus G\"{a}rtner and Richard A. Norte and Simon Gr\"{o}blacher},
journal = {Optica},
number = {2},
pages = {170--176},
publisher = {Optica Publishing Group},
title = {Coherent mechanical noise cancellation and cooperativity competition in optomechanical arrays},
volume = {9},
month = {Feb},
year = {2022},
url = {https://opg.optica.org/optica/abstract.cfm?URI=optica-9-2-170},
doi = {10.1364/OPTICA.446434},
}

@article{Kippenburg2007,
       author = {{Kippenberg}, Tobias J. and {Vahala}, Kerry J.},
        title = "{Cavity Opto-Mechanics}",
      journal = {Optics Express},
         year = 2007,
        month = dec,
       volume = {15},
       number = {25},
        pages = {17172},
          doi = {10.1364/OE.15.017172},
archivePrefix = {arXiv},
       eprint = {0712.1618},
 primaryClass = {physics.optics},
}

@article{Metcalfe2014,
    author = {Metcalfe, Michael},
    title = {Applications of cavity optomechanics},
    journal = {Applied Physics Reviews},
    volume = {1},
    number = {3},
    pages = {031105},
    year = {2014},
    month = {09},
    issn = {1931-9401},
    doi = {10.1063/1.4896029},
    url = {https://doi.org/10.1063/1.4896029},
}

@article{Li2021,
author = {Li, Bei-Bei and Ou, Lingfeng and Lei, Yuechen and Liu, Yong-Chun},
title = {Cavity optomechanical sensing},
journal = {Nanophotonics},
volume = {10},
number = {11},
pages = {2799-2832},
doi = {https://doi.org/10.1515/nanoph-2021-0256},
year = {2021}
}

@ARTICLE{Millen2020,
       author = {{Millen}, James and {Monteiro}, Tania S. and {Pettit}, Robert and {Vamivakas}, A. Nick},
        title = "{Optomechanics with levitated particles}",
      journal = {Reports on Progress in Physics},
         year = 2020,
        month = feb,
       volume = {83},
       number = {2},
          eid = {026401},
        pages = {026401},
          doi = {10.1088/1361-6633/ab6100},
archivePrefix = {arXiv},
       eprint = {1907.08198},
 primaryClass = {physics.optics},
}

@article{Ranjit:2016dlq,
    author = "Ranjit, Gambhir and Cunningham, Mark and Casey, Kirsten and Geraci, Andrew A.",
    title = "{Zeptonewton force sensing with nanospheres in an optical lattice}",
    eprint = "1603.02122",
    archivePrefix = "arXiv",
    primaryClass = "physics.optics",
    doi = "10.1103/PhysRevA.93.053801",
    journal = "Phys. Rev. A",
    volume = "93",
    number = "5",
    pages = "053801",
    year = "2016"
}

@ARTICLE{Marquardt2009,
       author = {{Marquardt}, Florian and {Girvin}, S.~M.},
        title = "{Optomechanics (a brief review)}",
      journal = {arXiv e-prints},
         year = 2009,
        month = may,
          eid = {arXiv:0905.0566},
archivePrefix = {arXiv},
       eprint = {0905.0566},
 primaryClass = {cond-mat.mes-hall},
}

@article{Maurer:2022nvj,
    author = "Maurer, Patrick and Gonzalez-Ballestero, Carlos and Romero-Isart, Oriol",
    title = "{Quantum theory of light interaction with a Lorenz-Mie particle: Optical detection and three-dimensional ground-state cooling}",
    eprint = "2212.04838",
    archivePrefix = "arXiv",
    primaryClass = "physics.optics",
    doi = "10.1103/PhysRevA.108.033714",
    journal = "Phys. Rev. A",
    volume = "108",
    number = "3",
    pages = "033714",
    year = "2023",
    note = "[Erratum: Phys.Rev.A 109, 019901 (2024)]"
}

@article{Maurer:2021zgx,
    author = "Maurer, Patrick and Gonzalez-Ballestero, Carlos and Romero-Isart, Oriol",
    title = "{Quantum electrodynamics with a nonmoving dielectric sphere: quantizing Lorenz{\textendash}Mie scattering}",
    eprint = "2106.07975",
    archivePrefix = "arXiv",
    primaryClass = "quant-ph",
    doi = "10.1364/JOSAB.498540",
    journal = "J. Opt. Soc. Am. B",
    volume = "40",
    number = "12",
    pages = "3137--3155",
    year = "2023"
}

@article{Reigue:2022thv,
    author = "Reigue, Antoine and Fogliano, Francesco and Heringlake, Philip and Mercier de L{\'e}pinay, Laure and Besga, Benjamin and Reichel, Jakob and Pigeau, Benjamin and Arcizet, Olivier",
    title = "{Cavity Nano-Optomechanics with Suspended Subwavelength-Sized Nanowires}",
    eprint = "2212.02860",
    archivePrefix = "arXiv",
    primaryClass = "quant-ph",
    doi = "10.1103/PhysRevApplied.20.014025",
    journal = "Phys. Rev. Applied",
    volume = "20",
    number = "1",
    pages = "014025",
    year = "2023"
}

@ARTICLE{Komori2020,
       author = {{Komori}, Kentaro and {Enomoto}, Yutaro and {Ooi}, Ching Pin and {Miyazaki}, Yuki and {Matsumoto}, Nobuyuki and {Sudhir}, Vivishek and {Michimura}, Yuta and {Ando}, Masaki},
        title = "{Attonewton-meter torque sensing with a macroscopic optomechanical torsion pendulum}",
      journal = {\pra},
         year = 2020,
        month = jan,
       volume = {101},
       number = {1},
          eid = {011802},
        pages = {011802},
          doi = {10.1103/PhysRevA.101.011802},
archivePrefix = {arXiv},
       eprint = {1907.13139},
 primaryClass = {quant-ph},
}

@ARTICLE{Gavartin2012,
       author = {{Gavartin}, E. and {Verlot}, P. and {Kippenberg}, T.~J.},
        title = "{A hybrid on-chip optomechanical transducer for ultrasensitive force measurements}",
      journal = {Nature Nanotechnology},
         year = 2012,
        month = aug,
       volume = {7},
       number = {8},
        pages = {509-514},
          doi = {10.1038/nnano.2012.97},
archivePrefix = {arXiv},
       eprint = {1112.0797},
 primaryClass = {physics.optics},
}

@article{Zhou2021,
author = {Feng Zhou and Yiliang Bao and Ramgopal Madugani and David A. Long and Jason J. Gorman and Thomas W. LeBrun},
journal = {Optica},
number = {3},
pages = {350--356},
publisher = {Optica Publishing Group},
title = {Broadband thermomechanically limited sensing with an optomechanical accelerometer},
volume = {8},
month = {Mar},
year = {2021},
url = {https://opg.optica.org/optica/abstract.cfm?URI=optica-8-3-350},
doi = {10.1364/OPTICA.413117},
}

@article{Liang2023,
title = {Yoctonewton force detection based on optically levitated oscillator},
journal = {Fundamental Research},
volume = {3},
number = {1},
pages = {57-62},
year = {2023},
issn = {2667-3258},
doi = {https://doi.org/10.1016/j.fmre.2022.09.021},
url = {https://www.sciencedirect.com/science/article/pii/S2667325822003879},
author = {Tao Liang and Shaochong Zhu and Peitong He and Zhiming Chen and Yingying Wang and Cuihong Li and Zhenhai Fu and Xiaowen Gao and Xinfan Chen and Nan Li and Qi Zhu and Huizhu Hu},
}

@article{Ahn:2019cbu,
    author = "Ahn, Jonghoon and Xu, Zhujing and Bang, Jaehoon and Ju, Peng and Gao, Xingyu and Li, Tongcang",
    title = "{Ultrasensitive torque detection with an optically levitated nanorotor}",
    eprint = "1908.03453",
    archivePrefix = "arXiv",
    primaryClass = "physics.app-ph",
    doi = "10.1038/s41565-019-0605-9",
    journal = "Nature Nanotech.",
    volume = "15",
    pages = "89",
    year = "2020"
}

@article{Ju:2023egl,
    author = "Ju, Peng and Jin, Yuanbin and Shen, Kunhong and Duan, Yao and Xu, Zhujing and Gao, Xingyu and Ni, Xinjie and Li, Tongcang",
    title = "{Near-field GHz rotation and sensing with an optically levitated nanodumbbell}",
    eprint = "2301.10868",
    archivePrefix = "arXiv",
    primaryClass = "quant-ph",
    doi = "10.1021/acs.nanolett.3c02442",
    journal = "Nano Lett.",
    volume = "23",
    pages = "10157",
    year = "2023"
}

@article{Guha2020,
  title = {Force Sensing with an Optomechanical Self-Oscillator},
  author = {Guha, Biswarup and Allain, Pierre Etienne and Lema\^{\i}tre, Aristide and Leo, Giuseppe and Favero, Ivan},
  journal = {Phys. Rev. Appl.},
  volume = {14},
  issue = {2},
  pages = {024079},
  numpages = {11},
  year = {2020},
  month = {Aug},
  publisher = {American Physical Society},
  doi = {10.1103/PhysRevApplied.14.024079},
  url = {https://link.aps.org/doi/10.1103/PhysRevApplied.14.024079}
}

@article{FormalismPaper,
    author = "Laeuger, Andrew and Aggarwal, Nancy",
    title = "{Mathematical framework for simulation of non-perturbative cavity optomechanics with plate-like dielectrics}",
    eprint = "2610.17475",
    archivePrefix = "arXiv",
    primaryClass = "gr-qc",
    month = "10",
    year = "2026",
    journal = {}
}

@article{ForceLetter,
      title={The optical force on a dielectric slab}, 
      author={Andrew Laeuger and Kevin Kuns and Yanbei Chen and Nancy Aggarwal},
      year={2026},
      eprint={2610.27254},
      archivePrefix={arXiv},
      primaryClass={physics.optics},
      journal={}
}

@article{Muller:2015idf,
    author = {M{\"u}ller, Tina and Reinhardt, Christoph and Sankey, Jack C.},
    title = "{Enhanced Optomechanical Levitation of Minimally Supported Dielectrics}",
    eprint = "1412.7733",
    archivePrefix = "arXiv",
    primaryClass = "quant-ph",
    doi = "10.1103/PhysRevA.91.053849",
    journal = "Phys. Rev. A",
    volume = "91",
    pages = "053849",
    year = "2015"
}

@article{Dorsel1983,
  title = {Optical Bistability and Mirror Confinement Induced by Radiation Pressure},
  author = {Dorsel, A. and McCullen, J. D. and Meystre, P. and Vignes, E. and Walther, H.},
  journal = {Phys. Rev. Lett.},
  volume = {51},
  issue = {17},
  pages = {1550--1553},
  numpages = {0},
  year = {1983},
  month = {Oct},
  publisher = {American Physical Society},
  doi = {10.1103/PhysRevLett.51.1550},
  url = {https://link.aps.org/doi/10.1103/PhysRevLett.51.1550}
}

@misc{Supplemental,
  note = {See Supplemental Material at [URL will be inserted by publisher].}
}

\onecolumngrid
\newpage
\section{Supplemental Material}

\subsection{Avoided crossings for highly reflective sensors}
The creation of two highly coupled subcavities by the insertion of the dielectric sensor is a familiar behavior in membrane-in-the-middle studies \cite{Thompson:2007hno,jayich2008dispersive,Xu:2013hgm,xu2015quantum,stambaugh2015membrane,Dumont:2019ziy,DumontThesis,Yao:2024tpc,deJong2020}.
The distinction here is that the sensor is freely levitated and only moderately reflective. The field reorganization is therefore inseparable from the mechanical equilibrium rather than being a small correction to a fixed optical mode. The static optical response of a membrane in a cavity was solved exactly in one dimension in Ref.~\cite{jayich2008dispersive}, where the reflectivity was varied at fixed, unphysically small thickness (0.01 nm) by changing the refractive index; here the thickness is varied at fixed $n$, which makes the $\lambda/2n$ periodicity explicit. Empty-cavity mode expansions have also been compared against a one-dimensional transfer-matrix solution for a high-index slab \cite{Muller:2015idf}.

\begin{figure}[h]
    \centering
    \includegraphics[width=0.75\linewidth]{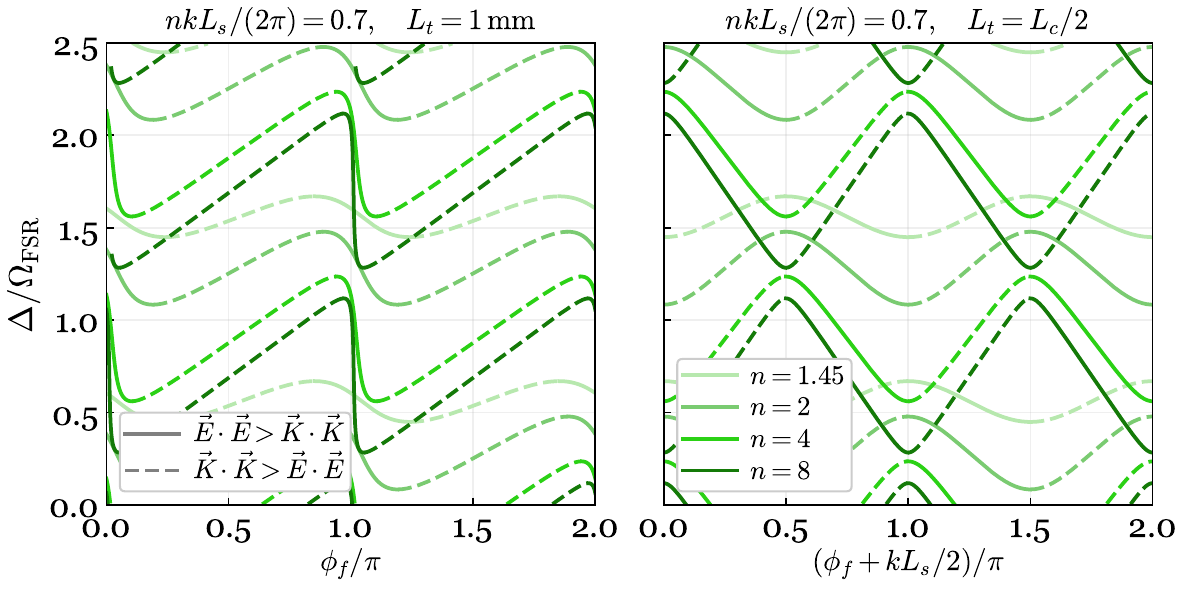}
    \includegraphics[width=0.8\linewidth]{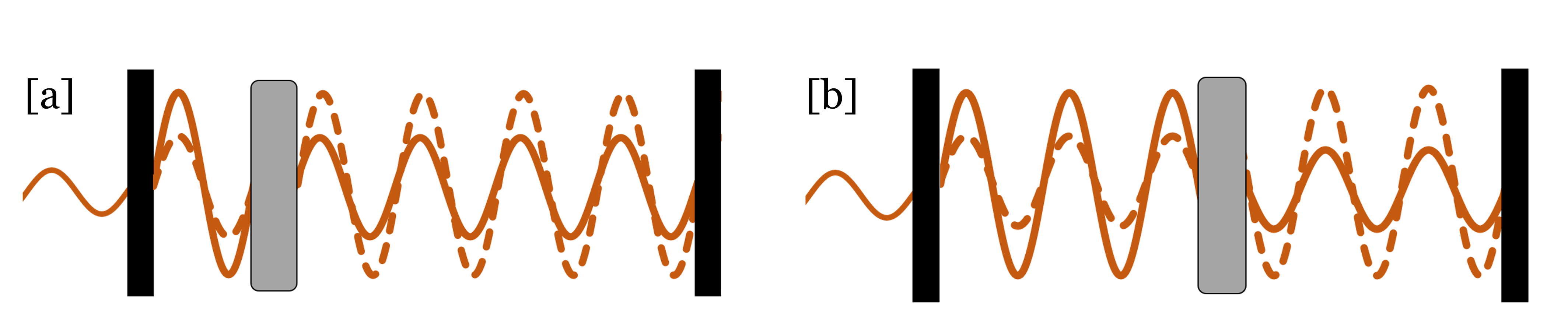}
    \caption{Optical resonances of the cavity-plus-sensor system as functions of sensor position and detuning, with $L_s=0.7\lambda/n$ and the macroscopic position set to either [a] near the input mirror or [b] the center of the cavity. Solid lines denote resonances led by Subcavity 1 and dashed lines those led by Subcavity 2. Increasing $n$ increases the sensor reflectivity and makes the avoided-crossing structure progressively more pronounced. The finite sensor thickness additionally vertically shifts the resonance bands through optical phase accumulation in the dielectric.}
    \label{fig:avoided}
\end{figure}

Nevertheless, we can reinforce this correspondence to the membrane literature: by increasing the refractive index of the sensor, we can continuously connect the resonance structures observed in this system to the recognizable avoided crossings produced by high reflectivity membranes. Figure~\ref{fig:avoided} shows the optical resonances as a function of sensor position and detuning for several values of $n$, placing the sensor either near the input mirror ($L_t=1$mm) or at the center of the cavity. We also include two sketches of the rough field profiles when the resonance is led by either Subcavity 1 or Subcavity 2.

As $n$ increases, the sensor becomes more reflective and the two cavities become increasingly decoupled, making the familiar avoided-crossing structure more pronounced. For a fixed phase accumulated over the sensor $nkL_s$, the resonance bands also shift vertically (compare to Figure 1 of \cite{Dumont:2019ziy}, for example) because the dielectric replaces an equivalent length of vacuum and accumulates additional optical phase.

\subsection{Convergence of optical modes via cavity locking}
In the main text, we described a ``locking" procedure by which we simultaneously maximize the transmitted power and keep the sensor at a stable mechanical equilibrium. We then demonstrated a number of consequences of implementing such a scheme. Usually, the more practical solution is to employ a Pound-Drever-Hall locking procedure which keeps the cavity length slightly off this resonance \cite{PDHlocking}, but for the sake of minimizing the number of free parameters, we assume the true maximum can be attained. 

To locate the parameters of the ``lock" in this calculation, we iteratively swap between maximizing the transmitted power over $\phi_{e,\text{trap}}$ and computing the new trap location $\phi_{f,\text{trap}}$ until both values find an equilibrium within a predetermined tolerance. We find that up to trivial additions of $\pm\pi$ to either phase, only a single point $(\phi_{f,\text{lock}},\phi_{e,\text{lock}})$ in the parameter space exists for a given $L_s$ which simultaneously solves Eq. \eqref{eq:trap} and provides a local maximum in the transmitted power, and our iterative solution method always finds this solution.

\begin{figure}[h]
    \centering
    \includegraphics[width=0.55\linewidth]{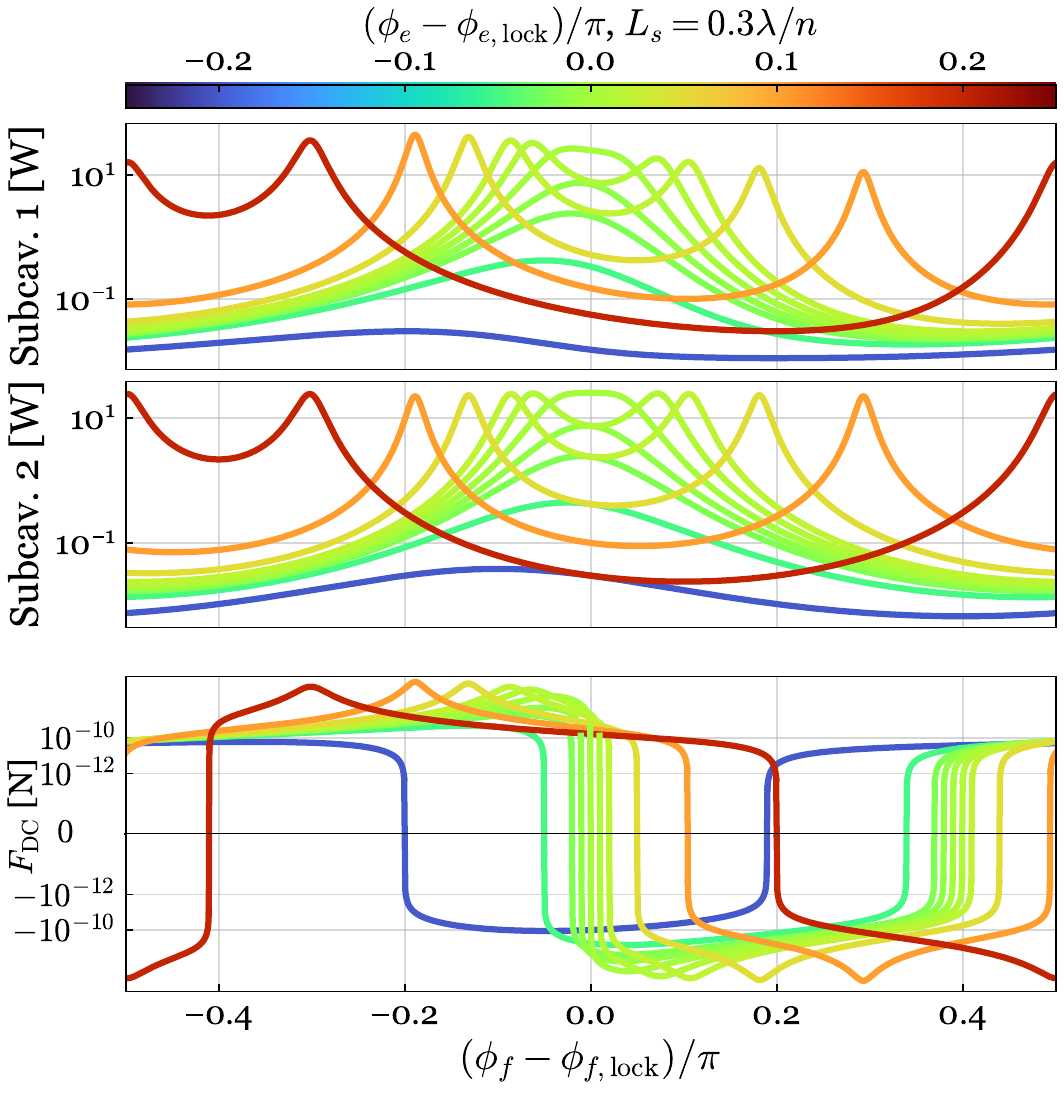}
    \caption{The subcavity powers and DC force on the sensor versus $\phi_f$ for various values of $\phi_e$ near its locked value $\phi_{e,\text{lock}}$, where we have fixed $nkL_s/2\pi=0.3$. The two distinct optical resonances which were identified for $\phi_e=0$ in Figure \ref{fig:subcavitylocked} begin to converge when $\phi_e$ gets within $\sim0.05\pi$ of $\phi_{e,\text{lock}}$, eventually forming a singly-peaked but non-Lorentzian resonance structure.}
    \label{fig:approachinglock}
\end{figure}

In Figure \ref{fig:approachinglock}, we demonstrate the convergence of the distinct optical resonances as $\phi_e$ approaches the locked solution. For $\phi_e>\phi_{e,\text{lock}}$ (to a reasonable extent), the two resonant peaks are apparent, with their separation decreasing as $\phi_e$ approaches the lock. At $\phi_{e,\text{lock}}$, a single peak has formed, though it is clearly not described by a Lorentzian distribution. Interestingly, for $\phi_e<\phi_{e,\text{lock}}$, the single peak structure remains, rather than the two resonances splitting once again; however, the peak transmitted power (which is proportional to the power circulating in Subcavity 2) purely decreases, so the algorithm for identifying the lock does not settle on these solutions.

The heavily-flattened power in Subcavity 2 near its peak makes a locking procedure based purely on measurements of the transmitted power impractical. The power in Subcavity 1 does show a linear response to end mirror displacement, however, so perhaps a routine which measures the reflected power from the cavity or relies on a hybrid measurement of the reflected and transmitted power could enable locking to this useful configuration.

\begin{figure}[h]
    \centering
    \includegraphics[width=0.6\linewidth]{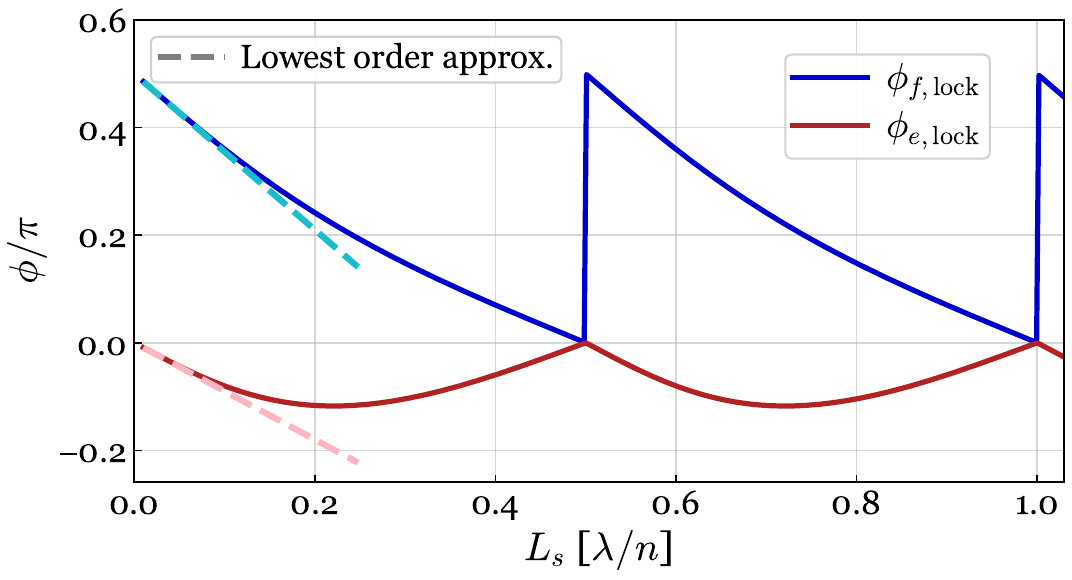}
    \caption{The solutions to the simultaneous cavity locking conditions $\phi_{f,\text{lock}}$ and $\phi_{e,\text{lock}}$ as a function of $L_s$, along with the lowest order perturbative result in Eq. \eqref{eq:lockpert}. The dashed lines match the exact solution for small $L_s$, but once again, the perturbative results depart from the exact solution by $L_s\sim\lambda/(10n)$.}
    \label{fig:lockpositions}
\end{figure}

We can analytically expand the fields and forces at lowest order in $kL_s$ to predict the behavior of the trap site and end mirror locations versus the sensor thickness. Doing so predicts that in the $L_s\rightarrow0$ limit,
\begin{equation}
    \frac{d\phi_{f,\text{lock}}}{d(kL_s)}=-n^2/2,\quad \frac{d\phi_{e,\text{lock}}}{d(kL_s)}=n(1-n).
    \label{eq:lockpert}
\end{equation}
In Figure \ref{fig:lockpositions}, we compute the exact locked cavity solutions and compare them to the lowest-order prediction above. Computing $\phi_{f/e,\text{lock}}$ to linear order in $kL_s$ in this way matches the observed numerical results well up to the familiar threshold of $L_s\sim \lambda/10n$, after which higher order corrections become increasingly important.

We also observe that the locked trap site $\phi_{f,\text{lock}}$ always lies within a range of width $\pi/2$, rather than the scale of $\pi$ over which the fields are periodic. The reason for this is straightforward: no matter the value of $\phi_e$, there are always exactly two points where the DC force goes to zero. However, only one of them has $\partial_{\phi_f}F_\text{DC}<0$, which is needed to generate a stable trap. At the points where $\phi_{f,\text{lock}}$ wraps back to $\pi/2$, the force is zero everywhere; as the thickness approaches these values, the two zero-crossings of $F_\text{DC}$ come to be separated by $\pi/2$. If the track were followed continuously, the stable site for $L_s<\lambda/2n$ would become the unstable site after crossing $L_s=\lambda/2n$ (i.e., the sign of $\partial_{\phi_f}F_\text{DC}$ flips at that point passing through $L_S=N\times\lambda/2n$). Thus the jumps are simply the sensor finding the stable equilibrium (which retains the same $\lambda/2n$ periodicity as the fields) after crossing the transparent points.

\subsection{Nonperturbative cavity frequency shift}

For a cavity with nodes of the electric field at its boundaries, the lowest order perturbative shift in the resonant frequency due to the insertion of the thin dielectric (Eq. \eqref{eq:pertshift} reduces to
\begin{equation}
\frac{\delta\omega_c}{\omega_c}
\approx
-\frac{n^2-1}{4kL_c}
\left\{2kL_s+\sin(2\phi_f)-\sin(2\phi_f+2kL_s)\right\}.
\label{eq:pertshift2}
\end{equation}
Our definition of $\phi_e$ is such that for an empty cavity, setting $\phi_e=0$ is equivalent to making a zero-detuned carrier resonant, i.e., $\omega_c=\omega_0$. We can also consider the next-to-lowest order perturbative frequency shift, which includes the modified electric field in the cavity upon insertion of the dielectric, $\vec{\mathbb{E}}_1$ \cite{Joannopoulos_Johnson_Winn_Meade_2008}:
\begin{equation}
    \frac{\delta\omega_c}{\omega_c}\approx-\frac12\frac{\int d^3{\vec r}\,' \delta\epsilon({\vec r}\,')\vec{ \mathbb{E}}_0({\vec r}\,')\cdot \vec{\mathbb{E}}_1({\vec r}\,')}{\int d^3{\vec r}\,' \vec{\mathbb{E}}_0({\vec r}\,')\cdot \vec{\mathbb{E}}_1({\vec r}\,')}.
    \label{eq: next lowest order frequency shift}
\end{equation}

\begin{figure}[h]
    \centering
    \includegraphics[width=0.6\linewidth]{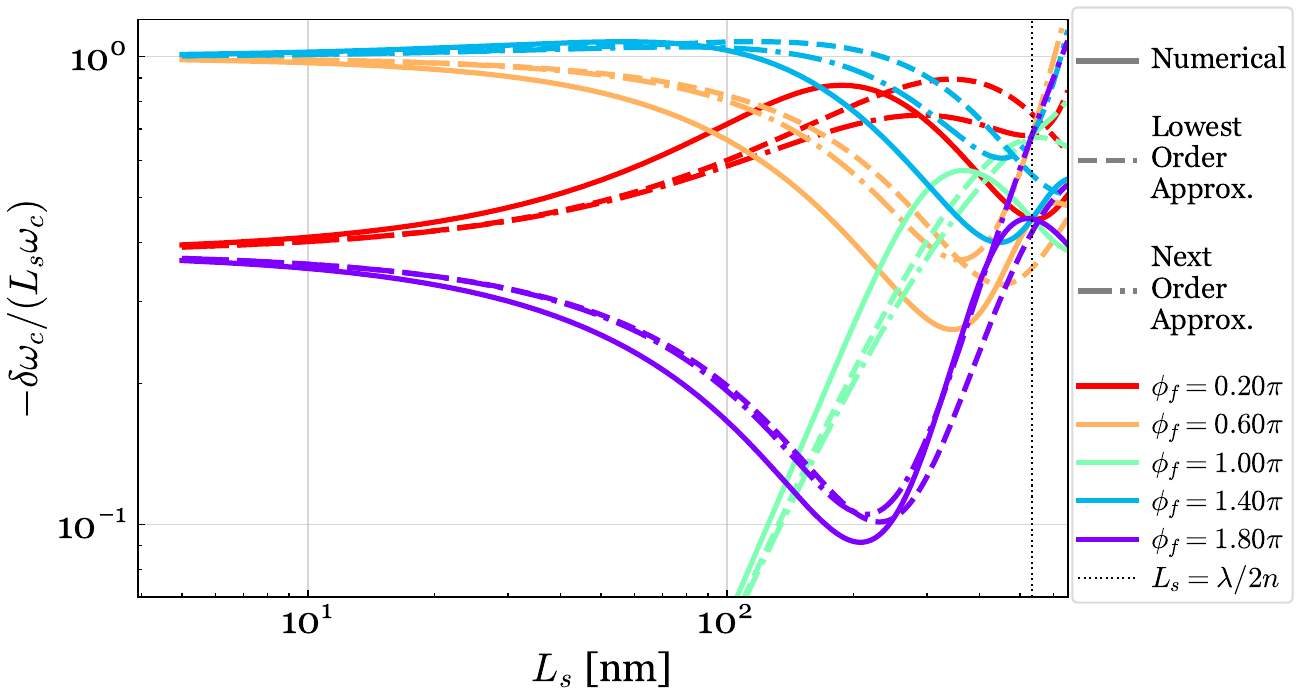}
    \caption{The cavity frequency shift due to the insertion of a thin dielectric, normalized by sensor thickness, for several sensor positions. The cavity length is kept fixed by setting $\phi_e=(n-1)kL_s$. Solid curves are obtained by maximizing the transmitted power in the exact cavity solution; dashed and dot-dashed curves use the lowest- and next-to-lowest-order perturbative expressions. The perturbative results have already departed from the exact solution by $L_s\sim\lambda/(10n)$.}
    \label{fig:deltaomega_c}
\end{figure}

In Figure \ref{fig:g0}, we demonstrated the departure of the perturbative expressions for the optomechanical coupling $g_0$, derived from these frequency shift equations, from the exact results at thicknesses as low as $L_s\sim\lambda/(10n)$. In Figure \ref{fig:deltaomega_c}, we demonstrate the same departure from perturbation theory, but for the frequency shift $\delta\omega_c$ itself. In order to make a proper comparison to the perturbative approaches, the cavity length must be kept fixed upon insertion of the dielectric, which we achieve by setting $\phi_e=(n-1)kL_s$. Much like the optomechanical coupling, $\delta\omega_c$ departs from both perturbative approximations in the vicinity of $\lambda/(10n)$. Meanwhile, the frequency shift with $\phi_f=\pi$ goes to zero in the $L_s\rightarrow0$ limit because the sensor sits at a node of the background standing wave, matching Eq. \eqref{eq:pertshift2}.

\subsection{Regaining the standard $\Delta=0$ split between heating and cooling}
\begin{figure}[h]
    \centering
    \includegraphics[width=0.495\linewidth]{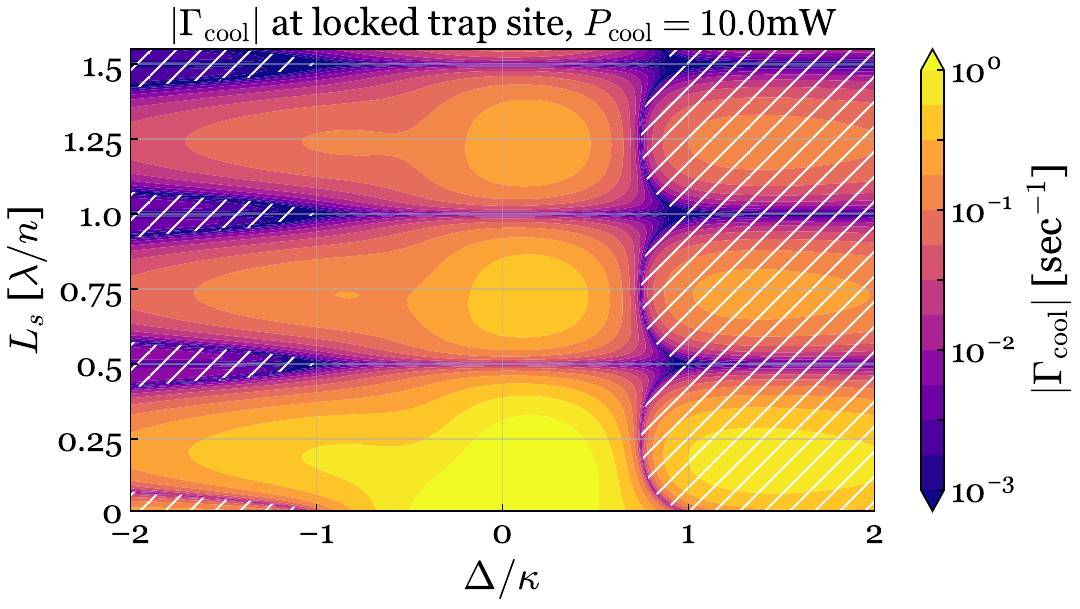}
    \includegraphics[width=0.485\linewidth]{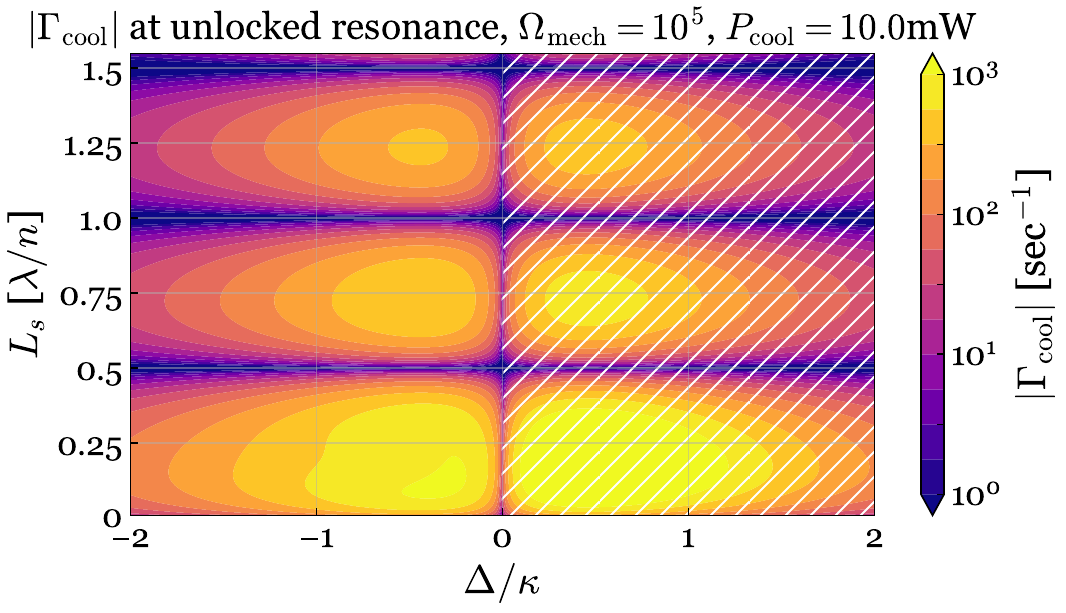}
    \caption{\textit{Left:} The induced cooling linewidth $\Gamma_\mathrm{cool}$ for a $10$-mW secondary laser as a function of its detuning, in units of the full-width half-maximum frequency $\kappa$ (with $\Delta=0$ at the transmission maximum), and the sensor thickness. The sensor and end-mirror positions are set by the cavity-locking condition for each $L_s$. Grey hatched regions indicate heating ($\Gamma_\mathrm{cool}<0$). Significant cooling persists at zero detuning as described in the main text. \textit{Right:} Equivalent to the left panel, except with $\phi_e=0$ and the sensor fixed at the point which maximizes the power in Subcavity 1 for zero detuning, assigning it a mechanical resonant frequency of $10^{5}\text{ s}^{-1}$. With the sensor at the resonance of a single distinct optical mode, the standard intuition holds where $\Delta=0$ marks the boundary between cooling and heating (grey hatches).}
    \label{fig:coolingunlocked}
\end{figure}

The appearance of significant cooling by a zero-detuned beam in Figure \ref{fig:cooling} is tied to the sensor equilibrium position lying off the peak in $\phi_f$, of a multimode optical resonance when the cavity is locked, even though the secondary beam is tuned to the transmission maximum ($\Delta=0$). In the left side of Figure \ref{fig:coolingunlocked}, we extend this result to a more general cooling approach, where a weak secondary beam of power $P_\text{cool}=10$mW and variable detuning enters the cavity. With the cavity locked and the sensor placed at the trap site, the structure of $\Gamma_\mathrm{cool}$ is indeed substantially more complicated than the standard intuition.
In the transparent-sensor limit, the cooling region is approximately bounded by $\sim\pm\kappa$. Generally, however, the cooling region is asymmetric, extending to approximately $-4.5\kappa$ on the red-detuned side and $+0.75\kappa$ on the blue-detuned side for $L_s=(2N+1)\lambda/(4n)$. Most notably, significant cooling is possible for a zero-detuned secondary beam. The detuned secondary also slightly changes the sensor's resonant frequency through the optical spring effect, but this effect is generally less remarkable than the induced cooling linewidth and thus omitted from our results.

We remarked in the main text that the standard intuition of the sign of $\Delta$ determining the cooling or heating could thus be recovered by generating a cavity configuration where the sensor indeed sits at the resonant peak of a single optical mode. We test this claim by setting $\phi_e=0$ so that the two resonances are clearly split and $\phi_f$ to maximize the power of a zero-detuned beam in Subcavity 1. Although this configuration does not correspond to a mechanical equilibrium for a levitated sensor, such a setup could be constructed by replacing the sensor and trap beam with a dielectric membrane which is mounted at that particular location and whose stiffness dictates the analogue to the trap resonant frequency. The results appear in the right side of Figure \ref{fig:coolingunlocked}, where we have imposed a mechanical resonant frequency of $\Omega_\text{mech}=10^5\text{ s}^{-1}$. As expected, the usual separation between cooling for detuning frequency $\Delta<0$ and heating for $\Delta>0$ is recovered.

\subsection{A caveat for mechanically supported intracavity membranes}

In Figure \ref{fig:coolingunlocked}, we invoked the possibility of using a mounted membrane to place the dielectric sensor at the peak of a single, isolated optical resonance, which recovers the standard intuition for the correlation between the sign of $\Delta$ and that of $\Gamma_\text{cool}$. Here, we expound on an important caveat for accurately modeling such membranes which does not present an issue for the freely levitated dielectrics we considered in the remainder of this work.

With the mount placed at some initial position $\phi_{f,0}$, the equilibrium position of the sensor $\phi_{f,\text{eq}}$ lies where the restoring force from the mount balances the optical force from the trapping beam -- notably, bistability can occur \cite{Dorsel1983}, so we choose the branch that would be found by slowly ramping the power up from zero. The subcavity powers and forces both vary substantially with the membrane position, so finding $\phi_{f,\text{eq}}$ requires the full position-dependent field solutions. Furthermore, the optical spring effect from the laser can modify the overall resonant frequency $\Omega_\text{eq}$.

\begin{figure}[h]
    \centering
    \includegraphics[width=0.95\linewidth]{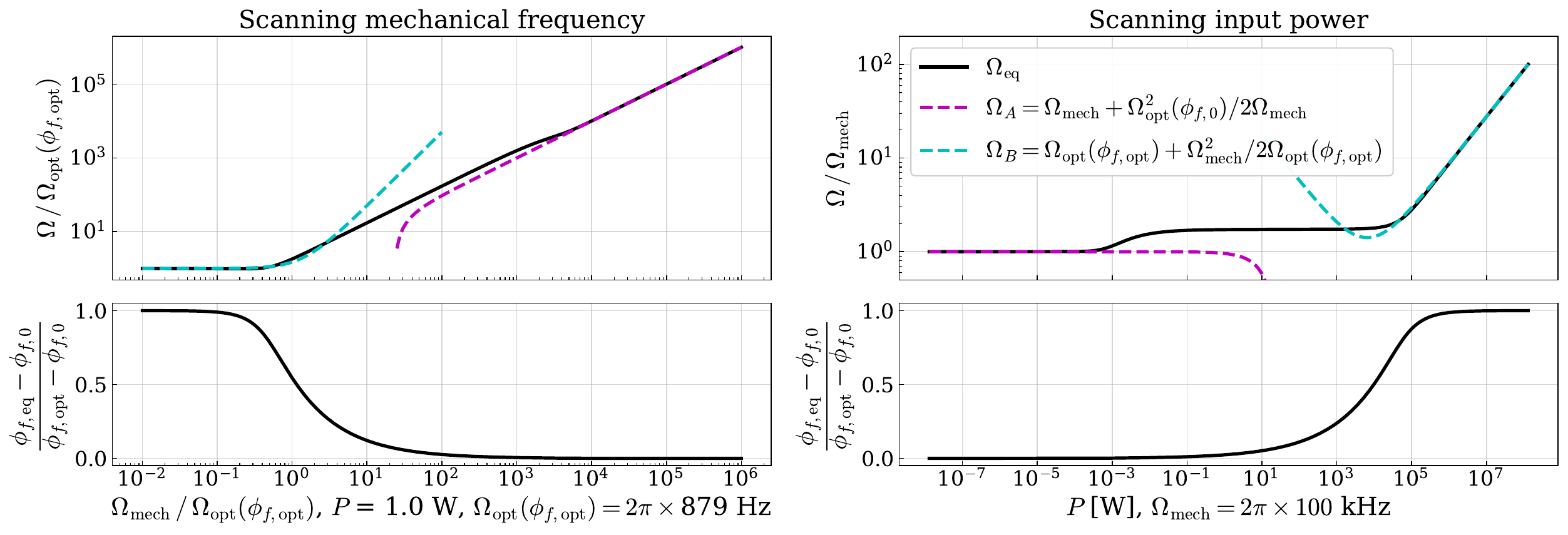}
    \caption{The equilibrium position $\phi_{f,\text{eq}}$ and resonant frequency $\Omega_\text{eq}$ for a $L_s=0.2\lambda/n$ membrane with mechanical resonant frequency $\Omega_\text{mech}$, mounted at $\phi_{f,0}$ which maximizes the power in Subcavity 1 for $\phi_e=0$ -- note that the cavity finesse has increased by a factor of ten to $\approx 770$. On the left, we scan the mechanical frequency while keeping $\Omega_\text{opt}$ fixed, and on the right, we vary the input laser power. The magenta and cyan dashed lines correspond to the perturbative frequency estimates in the stiff membrane and compliant membrane limits, respectively. Perturbations about either limit both fail to describe the true resonant frequency over a regime spanning four orders of magnitude in normalized frequency (eight in power). We emphasize that $\Omega_\text{opt}$, $\phi_{f,\text{opt}}$, $\Omega_\text{eq}$, and $\phi_{f,\text{eq}}$ are all computed using the forces derived from the full self-consistent cavity field equations.}
    \label{fig:membranetrapping}
\end{figure}

In the limit of a very stiff membrane, the membrane will only be slightly perturbed from its mounting point, and the strong mechanical restoring force results in $\Omega_\text{eq}$ varying only perturbatively from $\Omega_\text{mech}$. In the opposite limit, where the optical spring constant is much stronger, the equilibrium position should approach $\phi_{f,\text{opt}}$, the equilibrium with no mount, and $\Omega_\text{eq}$ is well-described by a perturbation about the frequency determined by the optical spring constant at that pure optical force equilibrium, $\Omega_\text{opt}(\phi_{f,\text{opt}})$. The intermediate regime, however, does not agree with either perturbative limit, as we demonstrate in Figure \ref{fig:membranetrapping} with a cavity whose finesse is increased to $\approx 770$.

We compute the exact equilibrium position and resonant frequency for a membrane of thickness $0.2\lambda/n$, initially positioned to maximize the power in Subcavity 1 for $\phi_e=0$ (i.e., the same position as in Figure \ref{fig:coolingunlocked}). On the left, we fix the power of the trap laser entering the cavity and vary the stiffness of the membrane mount. On the right, we fix the mechanical frequency and vary the input power, which rescales $\Omega_{\text{opt}}$. The magenta line $\Omega_A$ shows the perturbative frequency predicted in the stiff membrane limit -- notably, the perturbation is dictated by the optical spring constant at $\phi_{f,0}$, not $\phi_{f,\text{opt}}$. The cyan line $\Omega_B$ describes the strong optical trap limit instead. In the former limit, the equilibrium position is quite close to $\phi_{f,0}$, and in the latter, it approaches $\phi_{f,\text{opt}}$, as expected.

The perturbative predictions for the equilibrium resonant frequency match the exact result in the stiff membrane ($\Omega_\text{mech}\gg\Omega_\text{opt}$) and compliant membrane ($\Omega_\text{opt}\gg\Omega_\text{mech}$) limits; however, the stiff membrane prediction breaks down at just 1 mW of input power -- the increased finesse decreases the optical spring constant at $\phi_{f,\text{opt}}$, which lies between the two resonances with $\phi_e=0$, but increases the spring constant near the resonance at $\phi_{f,0}$. Meanwhile, the stiff optical spring perturbation $\Omega_B$ breaks down once the mechanical frequency becomes comparable to the free optical trap frequency, leaving a region spanning many orders of magnitude where neither $\Omega_A$ nor $\Omega_B$ agree with the exact, self-consistently computed resonant frequency.
We also find that as the strength of the mechanical restoring force grows, the equilibrium begins migrating away from $\phi_{f,\text{opt}}$ before $\Omega_\text{eq}$ departs meaningfully from the strong optical spring limit $\Omega_B$. These findings emphasize the need for self-consistent treatment of the complete optomechanical system: the true properties can deviate noticeably from those that would be predicted by assuming the membrane can be placed at any generic location, even when its stiffness well exceeds the pure optical spring constant.

\end{document}